\documentclass[11pt,a4paper]{article}
\usepackage{amsmath,amssymb}
\usepackage{epsfig,graphicx}

\begin{document}

\title{\bf On Consistent Lagrangian Quantization of Yang--Mills
Theories without Gribov Copies}
\author{A.~A.~Reshetnyak\footnote{{\bf e-mail}: reshet@ispms.tsc.ru}
\\
\small{\em Institute of Strength Physics and Materials Science} \\
\small{\em Siberian Branch of Russian Academy of Sciences} \\
\small{\em  634021, Tomsk,
Russia,}\\
\small{\em Tomsk State Pedagogical University} \\
\small{\em  634061, Tomsk,
Russia}}
\date{}
\maketitle

\begin{abstract}
We review the results of our research [A.A. Reshetnyak,  IJMPA  29
(2014) 1450184; P.Yu. Moshin, A.A. Reshetnyak,  Nucl. Phys. B 888
(2014)  92; P.Yu. Moshin, A.A. Reshetnyak,  Phys. Lett. B 739
(2014) 110; P.Yu. Moshin, A.A. Reshetnyak,
arXiv:1406.5086[hep-th]], devoted  to a consistent Lagrangian
quantization for gauge theories with soft BRST symmetry breaking,
in particular, for various descriptions of the Yang--Mills theory
without Gribov copies. The cited works rely on finite BRST
and BRST-antiBRST transformations, respectively, with
a singlet $\Lambda$ of nilpotent  and a doublet $\lambda_{a}$,
$a=1,2$, of anticommuting Grassmann parameters, both global and
field-dependent. It turns out that global finite BRST and
BRST-antiBRST transformations form a 1-parametric and a
2-parametric Abelian supergroup, respectively.  Explicit
superdeterminants corresponding to these changes of variables in
the partition function allow one to calculate precise changes of
the respective gauge-fixing functional. These facts provide the
basis for a proof of gauge independence of the corresponding path
integral under respective BRST and BRST-antiBRST transformations
and lead to the appearance of modified Ward identities. It is
shown that the gauge independence becomes restored for path
integrals with soft BRST and BRST-antiBRST symmetry breaking
terms. In this case, the form of transformation parameters is
found to induce a precise change of the gauge in the path
integral, thus connecting two arbitrary $R_{\xi }$-like gauges in
the average effective action. Finite field-dependent BRST-antiBRST
transformations are used to solve (perturbatively) the Gribov
problem in the Gribov--Zwanziger approach. A modification of the
path integral for theories with a gauge group, being consistent
with gauge invariance and providing a restriction of the
integration measure to the first Gribov region with a
non-vanishing Faddeev--Popov determinant, is suggested.

\end{abstract}

\section{Introduction} \label{introd}

It is well known that the electroweak and strong interactions are
described by the Standard Model, with the Quantum Chromodynamics
(QCD) as its constituent, and there are no experimental facts in
conflict with QCD. While the Standard Model has been justified by
the discovery of the Higgs boson, the problem of consistency in
QCD is far from its solution, especially in view of the
confinement phenomenon. The Lagrangian of QCD (and generally that
of the Standard Model) belongs to the class of non-Abelian gauge
theories~\cite{books1,books2,Bogolyubov,FaddeevSlavnov}
of Yang--Mills (YM) type.
Descendants of gauge invariance that emerge as one applies
the Faddeev--Popov trick \cite{FP} to the partition function
of Yang--Mills theories \cite{YM} are special supersymmetries,
known as the BRST symmetry \cite{BRST1,BRST2} and the BRST-antiBRST
symmetry \cite{aBRST1,aBRST2,aBRST4}. They provide a basis for
contemporary quantization methods applied to gauge theories \cite{books2, books}.
These symmetries are characterized by the presence of a Grassmann-odd
parameter $\mu$ and two Grassmann-odd scalar parameters
$(\mu,\bar{\mu})$, respectively. The latter parameters in the extended
schemes of generalized
Hamiltonian \cite{BLT1h} and Lagrangian \cite{BLT1,BLT2}
quantization (see \cite{Hull} as well) form
an $\mathrm{Sp}\left(  2\right) $-doublet  $(\mu,\bar{\mu})\equiv(\mu_{1}%
,\mu_{2})=\mu_{a}$, whereas the former parameter is used
in the generalized canonical \cite{BFV}, \cite{Henneaux1}
and field-antifield \cite{BV} quantization methods.
When considered either as constants or as field-dependent
functionals, these infinitesimal odd-valued parameters
can be used, respectively, to obtain the Ward identities
and to establish the gauge-independence of the partition
function in the path integral approach.

On the other hand, due to the well-known Gribov problem \cite{Gribov}, the
covariant quantization of YM and gravity theories on a basis of the
FP procedure cannot be  realized correctly for the entire spectrum
of the momenta distribution in the deep infra-red region for gauge fields
once the gauge condition has been imposed using differentiation \cite{Singer},
since there remains an infinitely large number of discrete gauge copies
after gauge-fixing, contrary to the case of Quantum Electrodynamics
(with an Abelian gauge symmetry). This implies that a non-Abelian addition
to the quadratic part of a Lagrangian turns a free well-defined
partition function (of Gauss-type) to one for an interacting theory,
which does not meet the requirement of positive definiteness
for density of the distribution function, due to an infinite number
of zero eigenvalues for the FP  matrix. Gribov has studied YM theories
in the Coulomb gauge and suggested a restriction of the domain
of functional integration for gauge fields to the so-called first
Gribov region, which has been effectively incorporated into
the functional measure as the Heaviside $\Theta$-function
($\mathcal{V}(1-\sigma(0,A))$ for vanishing momenta $k$
in the notations  \cite{Gribov}), thus realizing the ``no-pole''
condition for the ghost propagator and a correct interpretation
of the partition function.

There are other means of solving the Gribov problem: first,
the Gribov--Zwanziger (GZ) procedure \cite{Zwanziger},
where the mentioned $\Theta$-function, due to certain
non-perturbative arguments, such as a replacement
of the $\Theta$-function by the $\delta$-function
and the hypothesis \cite{Zwanziger} ``the equivalence
of the microcanonical and canonical ensembles in classical
statistical mechanics is valid here, so that it is correct
to replace the $\delta$-function by the Boltzmann factor'',
may be applied to the Landau and Coulomb \cite{GZ2}
gauges with a hermitian FP operator and a special addition
to the standard FP action, known as the Gribov--Zwanziger
horizon functional \cite{Zwanziger,Zwanziger1}.
However, this
addition is not gauge-invariant and hence non-invariant
under the initial  BRST transformations.
Second, there is a procedure of imposing an algebraic
(rather than differential) gauge on auxiliary scalar fields
in a theory with the ghost and antighost fields considered
as classical gauge fields, which is non-perturbatively
equivalent to a Yang--Mills theory with the same gauge group,
but with the FP operator considered as part of the classical
Lagrangian \cite{Slavnoveq,Slavnoveq1,Slavnoveq2}.

In \cite{JM}, BRST transformations with a finite field-dependent
parameter (FFDBRST) for YM theories with the FP quantum action
have been first introduced by using a functional equation for the
corresponding infinitesimal parameter, so as to provide the path
integral with
such a change of variables that would allow one to relate the
quantum action given in a certain gauge to the one given in a
different gauge, however, without solving this equation in a general
setting, which has led to the appearance of numerous similar results
(see \cite{Upadhyay1,Upadhyaya2} and references therein).
The problem of establishing a relation of
the FP action in a certain gauge with the action in a different
gauge by using a change of variables corresponding to a FFDBRST
transformation has been generally solved in \cite{LL1},
thus providing an exact relation between a finite parameter
and a finite change of the gauge-fixing condition. In
particular, this result leads to the preservation of the number of
physical degrees of freedom in a given YM theory with respect to
FFDBRST transformations, which means the impossibility of relating
the YM theory to another theory, e.g., to GZ action
\cite{Zwanziger1} in the same configuration
space. Notice that the study of Gribov copies in YM theories
has not been  restricted to a certain gauge; see the use of the
covariant and maximal Abelian gauges, as well as the Landau and
Coulomb gauges in
\cite{sorellas, sorellas1, sorellas2, Gongyo, 0806.0348, Reshetnyak2}.

Notice that the solution of a similar problem for arbitrary
constrained dynamical systems in the generalized Hamiltonian
formalism \cite{BFV, Henneaux1} has been recently proposed in
\cite{BLThf}, whereas for general gauge theories (featuring
reducible gauge symmetries and/or an open gauge algebra) an exact
Jacobian generated by FFDBRST transformations in the path integral
given by the BV procedure \cite{BV,BVref} has been obtained in
\cite{Reshetnyak} (see \cite{BLTBV} as well) and gives
a positive solution of the consistency
of \emph{soft BRST symmetry breaking} \cite{llr1}.

In \cite{MRnew}, we consider  an extension of BRST-antiBRST
transformations to the case of finite (both global and
field-dependent) parameters in YM theories. In \cite{MRnew2, MRnew3, MRnew1},
we have done the same for general gauge theories, by using the
Lagrangian and generalized Hamiltonian BRST-antiBRST quantization
methods; see also \cite{BLThfext}. In the present work, the origins
of finite BRST and BRST-antiBRST transformations are reviewed
in Sections~\ref{sec2} and \ref{sec4} respectively, and then
in Section~\ref{sec3} we use their properties to study their influence
on the quantum structure of YM theories and general gauge theories,
both with and without BRST(antiBRST) symmetry breaking terms, and
using the respective BRST and BRST-antiBRST settings,
including the cases of refined and standard GZ theories
in Section~\ref{sec5}.
A modification of the path integral in YM theories, which is consistent
with gauge invariance and which provides a restriction of the integration
measure to the first Gribov region with a non-vanishing FP determinant,
is suggested in Section~\ref{sec6}.

We use the condensed notation of DeWitt and the conventions of
\cite{Reshetnyak,MRnew}. Unless otherwise specified
by an arrow, derivatives with respect to the fields are taken
from the right, and those with respect to the corresponding
antifields are taken from the left. The raising and lowering
of $\mathrm{Sp}\left(  2\right)  $ indices,
$s^{a}=\varepsilon ^{ab}s_{b}$, $s_{a}=\varepsilon_{ab}s^{b}$, is
carried out using a constant antisymmetric metric tensor
$\varepsilon^{ab}$, $\varepsilon
^{ac}\varepsilon_{cb}=\delta_{b}^{a}$, subject to the
normalization $\varepsilon^{12}=1$.
The Grassmann  parity of a homogeneous quantity $B$
is denoted as $\varepsilon(B)$.

\section{Finite Field-dependent BRST Transformation and its Jacobian}\label{sec2}

An extended generating functional of Green's functions (GFGF) for a gauge theory defined in a total configuration
space $\mathcal{M}$ parameterized by fields $\phi^A$, $\varepsilon(\phi^A)=\varepsilon_A$, which, in the
formalism of BRST quantization \cite{BV}, contain the initial classical fields $A^i$, $\varepsilon(A^i)=\varepsilon_i$,
$i=1,\ldots,n$, the Nakanishi--Lautrup fields $B^{\alpha}$, $ \varepsilon(B^{\alpha})=\varepsilon_\alpha$,
$\alpha=1,\ldots,m<n$, and the pairs of ghost and antighost fields\footnote{As well as the towers of additional ghost,
antighost and Nakanishi--Lautrup fields, introduced according to the stage of reducibility
of a given theory \cite{BVref}.}
$C^{\alpha}$, $\bar{C}{}^{\alpha
}$, $\varepsilon(C^{\alpha})$ = $\varepsilon(\bar{C}{}^{\alpha})$ = $\varepsilon_\alpha+1$, is given by the rule
\begin{equation}
Z_\Psi(J,\phi^*)=\int d\phi\ \exp\left\{  \textstyle\frac{i}{\hbar}\left[  S_{\Psi}\left(  \phi, \phi^*\right)
+J_{A}\phi^{A}\right]  \right\}  \equiv\int\mathcal{I}_{\phi,\phi^*}^{\Psi}\exp\left\{
\textstyle\frac{i}{\hbar}J_{A}\phi^{A}\right\}  \;,\label{z(j)}%
\end{equation}
where  $\hbar, J_A,  \phi^*_A$ and $\Psi(\phi)$ are, respectively, the Planck constant, external sources
to $\phi^A$, antifields, $\varepsilon(J_A)=\varepsilon(\phi^*_A)+1=\varepsilon_A$ , and an admissible
Fermionic gauge-fixing functional $\Psi\left(  \phi\right) $.  The usual GFGF is
$Z_\Psi(J)=Z_\Psi(J,0)$ and the quantum action $S_{\Psi}\left(  \phi, \phi^*\right)  $ is given by
\begin{equation}
S_{\Psi}\left(  \phi, \phi^*\right)  = S\left(  \phi, \phi^* +\textstyle\frac{\delta \Psi}{\delta\phi}\right) , \  \mathrm{  where  } \    S\left(  \phi, 0\right) = \mathcal{S}_0(A),\label{action}%
\end{equation}
with the classical action $\mathcal{S}_{0}(  A)  $ invariant under infinitesimal gauge transformations
$\delta A^i = R^i_\alpha (A)\xi^\alpha $, $\varepsilon(R^i_\alpha )= \varepsilon_i + \varepsilon_\alpha $,
whose generators  $R^i_\alpha (A)$ form an algebra of gauge transformations,
\begin{align}
&  R_{\alpha,j}^{i}(A)R_{\beta}^{j}(A)-\left(  -1\right)  ^{\varepsilon
_{\alpha}\varepsilon_{\beta}}R_{\beta,j}^{i}(A)R_{\alpha}^{j}(A)=-R_{\gamma
}^{i}(A)F_{\alpha\beta}^{\gamma}\left(  A\right)  -S_{0,j}(A)M_{\alpha
\beta}^{ij}\left(  A\right)  \ ,\nonumber\\
&  \mathtt{for}\,\,F_{\alpha\beta}^{\gamma}=-\left(  -1\right)
^{\varepsilon
_{\alpha}\varepsilon_{\beta}}F_{\beta\alpha}^{\gamma}\ ,\ \
M_{\alpha\beta }^{ij}=-\left(  -1\right)
^{\varepsilon_{i}\varepsilon_{j}}M_{\alpha\beta
}^{ji}=-\left(  -1\right)  ^{\varepsilon_{\alpha}\varepsilon_{\beta}}%
M_{\beta\alpha}^{ij}\,.\label{gauge_alg}%
\end{align}
The bosonic functional $S=S(\phi, \phi^*)$, as well as the quantum action $S_\Psi$, satisfies
the master equation (in two equivalent forms)
з\begin{equation}\label{MEBV}
{\Delta}\exp\left\{\textstyle\frac{\imath}{\hbar}{ S}\right\} =0 \Longleftrightarrow
\textstyle\frac {1}{2} ({ S},{
S})\ =\  \imath\hbar\,{\Delta}{ S}\,,
\end{equation}
expressed in terms of an odd Poisson bracket $(\cdot,\cdot)$, also called antibracket, and in terms of an odd Laplacian
$\Delta = (-1)^{\varepsilon_A}\frac{\delta_l}{\delta\phi^A}\frac{\delta}{\delta\phi^*_A}$, defined in the field-antifield space \cite{BV}.
Finite (group) BRST transformations introduced in \cite{Reshetnyak} are invariance transformations for the integrand
$\mathcal{I}_{\phi,\phi^*}^{\Psi}$ with account taken of (\ref{MEBV}) for $S_\Psi$,
\begin{equation}\label{finBRST}
  \left(\phi^A, \phi^*_A\right) \to \left(\phi^{\prime A}, \phi^{\prime *}_A\right) =\left(\phi^A\exp\{ \overleftarrow{s}_e \Lambda\}, \phi^*_A\right)    \Longrightarrow \mathcal{I}_{\phi \exp\{ \overleftarrow{s}_e \Lambda\},\phi^*}^{\Psi}= \mathcal{I}_{\phi,\phi^*}^{\Psi},
\end{equation}
where the set $\{g(\Lambda)\}=\{\exp\{ \overleftarrow{s}_e \Lambda\}\} $ forms a one-parametric Lie supergroup with an odd parameter $\Lambda$,
despite the fact that the generator $\overleftarrow{s}_e = \frac{\overleftarrow{\delta}}{\delta\phi^A}\frac{{\delta} S_\Psi}{\delta\phi^*_A}$
of BRST transformations fails to be nilpotent, $\overleftarrow{s}_e^2 \ne 0$, due to the presence of $M_{\alpha\beta }^{ij} \ne 0$
in the gauge algebra relations (\ref{gauge_alg}) and also due to the presence of the operator  $\Delta$.
When the parameter $\Lambda$ is chosen as a  field-dependent functional $\Lambda(\phi,\phi^*)$ depending parametrically
on antifields,  the set of  $\{g(\Lambda)\}$ transforms into an non-Abelian supergroup. The superdeterminant of  a change of variables
corresponding to FFDBRST transformations (\ref{finBRST})  has been calculated  in \cite{Reshetnyak} and reads
\begin{equation}\label{sdet}
\hspace{-0.05em}\mathrm{Sdet}\left\Vert \frac{\delta \phi ^{ A}\exp\{ \overleftarrow{s}_e \Lambda(\phi,\phi^*) \} }{\delta \phi
{}^{B}}\right\Vert \ =\  \big(1+\Lambda \overleftarrow{s}_{e}\big)^{-1} \exp\{ \overleftarrow{s}_e \Lambda(\phi,\phi^*) \}\Big\{1 +  \bigl(\Delta S_{\Psi
}\bigr)\Lambda\Bigr\}
,
\end{equation}
with the notation $ \frac{{\delta} S_\Psi}{\delta\phi^*_A}\equiv S_{\Psi }^{A}$. For 
 constant $\Lambda$,
the Jacobian reduces to  $\mathrm{Sdet}\left\Vert \frac{\delta \phi ^{A} \exp\{ \overleftarrow{s}_e \Lambda(\phi,\phi^*)\}}{\delta \phi
{}^{B}}\right\Vert = \Big\{1 +  \bigl(\Delta S_{\Psi
}\bigr)\Lambda\Bigr\}$.
The requirement of gauge independence for a finite change of the gauge,\footnote{This change is inspired
 by infinitesimal FDBRST transformations \cite{BV, BVref} with $\Lambda\left(\phi\right) = - (\imath/\hbar )\delta \Psi$,
 for which the vacuum functional is gauge-independent under a variation of the gauge condition, $\Psi\to\Psi+\delta\Psi$:  $Z_{\Psi+\delta\Psi}(0,0)=Z_{\Psi}(0,0)$.}
$\Psi\to\Psi+\Delta_f \Psi$,
leads to the compensation equation \cite{Reshetnyak},
\begin{equation}\label{compeq}
  \mathcal{I}_{\phi \exp\{ \overleftarrow{s}_e \Lambda(\phi,\phi^*)\},\phi^*}^{\Psi}= \mathcal{I}_{\phi,\phi^*}^{\Psi+\Delta_f\Psi},
\end{equation}
being a functional equation for an unknown $\Lambda(\phi,\phi^*) $,
\begin{equation}\label{funccomprel}
  -i\hbar  \,\mbox{ln}\,\Big[\big(1+\Lambda \overleftarrow{s}_{e}\big)^{-1} \exp\{ \overleftarrow{s}_e \Lambda(\phi,\phi^*)\} \Big]  = \Big(\exp\Big\{ -[\Delta,\, {\Delta}_f \Psi] \Big\}-1\Big)S_\Psi
\end{equation}
which has been proven to have a solution \cite{Reshetnyak},
\begin{equation}\label{solfeq}
\Lambda\left(\phi,\phi ^{\ast }|{\Delta}_f \Psi\right)  = \Lambda\left({\Delta}_f \Psi\right) \,\, \ \mathrm{so\ \  that } \
\Lambda\left(\phi,\phi ^{\ast }|{\Delta}_f \Psi\right) = - (\imath/\hbar ){\Delta}_f \Psi + o({\Delta}_f \Psi)\,.
\end{equation}
This allows one to state the gauge independence of the vacuum functional under finite changes of the gauge Fermion.
Next, in theories having a closed algebra of rank $1$, i.e., $M_{\alpha\beta }^{ij} = 0$  in  (\ref{gauge_alg}) and
being such that $\Delta S_\Psi =0 $, provided that  $\overleftarrow{s}_{e}{}^2 =0$,  the Jacobian (\ref{sdet}),
the compensation equation (\ref{funccomprel}), and its solution (\ref{solfeq}) are reduced  to those
corresponding to $\hat{\Lambda }= \Lambda(\phi) $, namely,
\begin{eqnarray}
&& \mathrm{Sdet}\left\Vert \frac{\delta \phi ^{A} \exp\{ \overleftarrow{s}_e \hat{\Lambda}\}}{\delta \phi
{}^{B}}\right\Vert  \  =\  \big(1+\hat{\Lambda} \overleftarrow{s}_{e}\big)^{-1} ;  \ \ \ \ \  \imath\hbar\left\{\ln \big(1+ \hat{\Lambda}\overleftarrow{s}_e\big)\right\}\  =\  \bigl({\Delta}_f\Psi(\phi)\big)\overleftarrow{s}_e,
 \label{aollambda} \\
&&\hat{\Lambda}\  =\ {\Delta }_f\Psi (\phi
)\big\{\bigl({\Delta}_f\Psi(\phi)\big)\overleftarrow{s}_e\big\}%
^{-1}\Big[\exp \left\{\textstyle -\frac{\imath}{\hbar }\bigl({\Delta}_f\Psi(\phi)\big)\overleftarrow{s}_e\right\} -1\Big].\label{YMlambda}
\end{eqnarray}
Relations (\ref{aollambda}), (\ref{YMlambda})  are identical to those in YM theories \cite{LL1},
when restricted to the case of irreducible gauge theories, provided that $F_{\alpha\beta, i}^{\gamma}=0$.

\section{Gauge Dependence Problem and  Ward Identities for Gauge Theories with  Soft BRST Symmetry Breaking}\label{sec3}

A \emph{soft BRST symmetry breaking}  term is introduced into the gauge theory as an bosonic additions, $M=M(\phi,\phi^*)$,  to the quantum action, $S_\Psi$ \cite{llr1} thus determining the GFGF,
\begin{equation}
Z_{ M_\Psi,\Psi}(J,\phi^*)\ =\ \int d\phi\ \textstyle\exp\left\{  \frac{i}{\hbar}\left[  S_{\Psi}\left(  \phi, \phi^*\right)+M_\Psi(\phi,\phi^*)
+J\phi\right]  \right\}  \equiv\int\mathcal{I}_{\phi,\phi^*}^{M_\Psi, \Psi}\exp\left\{
\textstyle\frac{i}{\hbar}J\phi\right\}.  \label{z(jm)}%
\end{equation}
The BRST breaking term $M$ does not invariant with respect to the same   BRST transformations (\ref{finBRST}),  and may or not satisfies to the so-called \emph{soft BRST symmetry breaking equation} respectively for dimensional-like regularization when $\Delta M=0$, for local $M$ \cite{llr1}, and for more general regularization \cite{lrr}
\begin{equation}\label{sbrstbr}
  M \exp\{\overleftarrow{s}_e \Lambda\} = M +M_A (\phi^A\overleftarrow{s}_e )\Lambda \ne M \ \ \mathrm{and}\ \  (M,M)=0\ \ \mathrm{or}\ \ \textstyle {\Delta}\exp\left\{-\frac{\imath}{\hbar}{ M}\right\} =0,
\end{equation}
however, providing an existence of the vacuum functional $Z_{M_\Psi, \Psi}(0, 0)$.  As the consequence of the BRST breaking  the integrand $\mathcal{I}_{\phi,\phi^*}^{M_\Psi, \Psi}$ fails to be invariant for $J_A = 0$,
 \begin{equation}\label{nonbrstinv}
   \mathcal{I}_{\phi\exp\{\overleftarrow{s}_e \Lambda\},\phi^*}^{M_\Psi, \Psi} =  \mathcal{I}_{\phi\exp\{\overleftarrow{s}_e \Lambda\},\phi^*}^{0,\Psi}\exp\left\{
\textstyle\frac{i}{\hbar}M\exp\{\overleftarrow{s}_e \Lambda\}\right\}\stackrel{(\ref{sbrstbr})}{\ne} \mathcal{I}_{\phi,\phi^*}^{M_\Psi, \Psi}.
 \end{equation}
In spite of this fact, there is  a \emph{modified Ward identity} for $Z_{M_\Psi,\Psi}(J)$
which is easily obtained by making in (\ref{z(jm)}) a field-dependent BRST transformation
(\ref{finBRST}) and using the relations (\ref{solfeq})  and the expression
(\ref{sdet}) for the Jacobian:
\begin{equation}
\left\langle \left\{  1+\textstyle\frac{i}{\hbar}\left[J_{A}\phi^{A} + M_\Psi\right]\overleftarrow
{s}_e\Lambda\left({\Delta}_f \Psi\right)  \right\}  \left(  1+\Lambda\left({\Delta}_f \Psi\right)\overleftarrow
{s}_e\right)  {}^{-1} \exp\{ \overleftarrow{s}_e \Lambda\left({\Delta}_f \Psi\right) \}\right\rangle _{M,\Psi,J} =1\ , \label{mWIbvbr}%
\end{equation}
where the symbol \textquotedblleft$\langle\mathcal{A}\rangle_{M,\Psi,J}%
$\textquotedblright\ for a quantity $\mathcal{A}$ stands for
a source-antifield- dependent average expectation value with respect to $Z_{M_\Psi,\Psi}(J,\phi^*)$,
corresponding to the gauge-fixing $\Psi$. The modified Ward identity (derived then for Green's functions as well) depends on the field-dependent parameter $\Lambda\left({\Delta}_f \Psi\right)$  as the weight  functional, and therefore on the change of the gauge condition, $\Delta_f \Psi$.
Note, first, that (\ref{mWIbvbr}) for $M=0$ permits to obtain new form of modified Ward identity for the general gauge theories within BV quantization, second,   the Ward identity takes the form for a constant $\Lambda$,
\begin{eqnarray}
&\hspace{-1em}& \hspace{-1em}\left\langle \left[J_{A}\phi^{A}+M_{\Psi}\right]  \overleftarrow
{s}_e\right\rangle _{M,\Psi,J} =0   \Longleftrightarrow \hspace{-0.25em}\Big(\hspace{-0.15em}J_{A}\hspace{-0.15em}+ \hspace{-0.15em}M_{A}\big({\textstyle\frac{\hbar }{\imath}}{%
\textstyle\frac{\delta }{\delta J}},\phi ^{\ast }\big)\hspace{-0.15em}\Big)\hspace{-0.15em}\left(\hspace{-0.15em} \textstyle\frac{%
\hbar }{\imath}\frac{\delta }{\delta \phi _{A}^{\ast }}\hspace{-0.15em} -\hspace{-0.15em} M^{A\ast }\big(%
{\textstyle\frac{\hbar }{\imath}}{\textstyle\frac{\delta }{\delta J}},\phi
^{\ast }\hspace{-0.15em}\big)\hspace{-0.25em}\right)\hspace{-0.15em} Z_{M_\Psi,\Psi}\hspace{-0.15em}=0 ,  \label{mWIbvbr1}\\
&\hspace{-1em}&\hspace{-1em} \mathrm{for}\  M_{A}\big({\textstyle\frac{\hbar }{\imath}}{\textstyle\frac{\delta }{%
\delta J}},\phi ^{\ast }\big)\equiv \textstyle\frac{\delta M(\phi ,\phi ^{\ast })}{%
\delta \phi ^{A}}\Big|_{\phi \rightarrow \frac{\hbar }{\imath}\frac{\delta
}{\delta J}}\ \ \mathrm{and}\  \ M^{A\ast }\big({\textstyle\frac{\hbar }{%
\imath}}{\textstyle\frac{\delta }{\delta J}},\phi ^{\ast }\big)\equiv
\textstyle\frac{\delta M(\phi ,\phi ^{\ast })}{\delta \phi _{A}^{\ast }}\Big|_{\phi
\rightarrow \frac{\hbar }{\imath}\frac{\delta }{\delta J}},
\end{eqnarray}
 which is identical
with the Ward identity for $Z_{M_\Psi,\Psi}(J,\phi^*)$ in \cite{Reshetnyak, lrr}, whereas for $\Delta M=0$ one should to extract from the left-hand side of the latter identity, the term, $M_{A}\ M^{A\ast }=0$, deriving the identity from \cite{llr1}.

The identity (\ref{mWIbvbr}) together with equivalence theorem arguments \cite{equiv} implies an equation which describes the gauge dependence of $Z_{M_\Psi,\Psi}(J)$ for a finite
change of the gauge $\Psi\rightarrow \Psi'=\Psi+\Delta_f \Psi$, namely,%
\begin{align}
&Z_{M_{\Psi'},\Psi'}(J,\phi^*) - Z_{M_\Psi,\Psi}(J,\phi^*)   =\textstyle Z_{M_\Psi,\Psi}(J,\phi^*)\left\langle \frac{i}{\hbar}%
J_{A}\phi^{A} \overleftarrow{s}_e\Lambda\left(  \phi,\phi^*|-\Delta_f\Psi\right)
\right\rangle _{M,\Psi,J} \nonumber \\
&\qquad  =  (-1)^{\varepsilon_A} J_{A}\Lambda\left({%
\textstyle\frac{\hbar }{\imath}}{\textstyle\frac{\delta }{\delta J}},\phi^*|-{\Delta}_f \Psi\right)\left(\frac{\delta}{\delta\phi^*_A} -\frac{\imath}{\hbar } M^{A*}_\Psi \right) Z_{M_\Psi,\Psi}(J,\phi^*) ,\label{GDInewb}%
\end{align}
if the following representation for the soft BRST symmetry breaking term $M$ in the reference frame described by the gauge, $\Psi+\Delta_f\Psi$, holds
\begin{equation}\label{reprmpsi+}
  M_{\Psi+\Delta_f\Psi}(\phi,\phi^*)\  =\  M_\Psi(\phi,\phi^*)\exp\{ \overleftarrow{s}_e \Lambda\left({\Delta}_f \Psi\right) \}\  =\  M_\Psi(\phi,\phi^*)\big[1+ \overleftarrow{s}_e \Lambda\left({\Delta}_f \Psi\right) \big].
\end{equation}
We obtain the main result of .\cite{Reshetnyak} that on the extremals $J=0$ finite change of the GFGF for the gauge theory with (soft) BRST symmetry breaking vanishes therefore preserving the gauge independence property in case $M\ne 0$ providing that the form of BRST symmetry breaking term $M$ transforms under change of the gauge: $\Psi\to\Psi+\Delta_f\Psi$ by the rule (\ref{reprmpsi+}). The equation (\ref{reprmpsi+}) determines the rule of transformation of any quantity under change of the gauge.

For the effective action (generating functional of vertex Green's functions), $\Gamma_{M_\Psi,\Psi}$ = \newline $\Gamma_{M_\Psi,\Psi}(\phi,\phi^*)$,  obtained via Legendre transformation of $\ln Z_{M_\Psi,\Psi}$   with respect to $J_A$,
\begin{equation}\label{eam}
\Gamma_{M_\Psi,\Psi}(J,\phi^*) = \frac{\hbar}{\imath}\ln Z_{M_\Psi,\Psi}(J,\phi^*) - J\phi, \ \ \mathrm{with}\ \  \phi^A =
\frac{\hbar}{\imath}\frac{\delta \ln Z_{M_\Psi,\Psi}}{\delta J_A}, \quad
\frac{\delta\Gamma_{M_\Psi,\Psi}}{\delta\phi^A}=-J_A.
\end{equation}
the  Ward identity (\ref{mWIbvbr1}) takes the form \cite{llr1, Reshetnyak} in terms of antibracket and operatorial fields ${\widehat\phi}{}^A$:
\begin{eqnarray}
&\hspace{-0.5em}&\hspace{-0.5em} {\textstyle\frac{1}{2}}(\Gamma _{M},\Gamma _{M})\ =\frac{\delta \Gamma _{M}%
}{\delta \Phi ^{A}}{\widehat{M}}^{A\ast }+{\widehat{M}}_{A}\frac{\delta
\Gamma _{M}}{\delta \Phi _{A}^{\ast }}-{\widehat{M}}_{A}{\widehat{M}}^{A\ast
}\ , \label{mWIG}\\
\label{MAG}
&\hspace{-0.7em}& \hspace{-0.9em}\mathrm{for} \  {\widehat{M}}_{A}\ \equiv \ \textstyle\frac{\delta M(\phi ,\phi ^{\ast })}{\delta \phi
^{A}}\Big|_{\phi \rightarrow \widehat{\phi }}, \qquad  {%
\widehat{M}}^{A\ast }\ \equiv \ \frac{\delta M(\phi ,\phi ^{\ast })}{\delta
\phi _{A}^{\ast }}\Big|_{\phi \rightarrow \widehat{\phi }},\\%
&\hspace{-0.7em}&\hspace{-0.9em}\textstyle \mathrm{and}\; {\widehat\phi}{}^A \hspace{-0.1em}= \phi^A \hspace{-0.1em}+ \hspace{-0.1em} \imath\hbar\,(\Gamma^{''-1}_{M,\Psi})^{AB}
\frac{\delta_l}{\delta\phi^B},\ (\Gamma^{''}_{M,\Psi})_{AB}\hspace{-0.1em} = \hspace{-0.1em}\frac{\delta_l}{\delta\phi^A}
\Big(\frac{\delta\Gamma_{M,\Psi}}{\delta\phi^B}\Big): \;
(\Gamma^{''-1}_{M,\Psi})^{AC}(\Gamma^{''}_{M,\Psi})_{CB}\hspace{-0.1em} = \hspace{-0.1em} \delta^A_{\; B}. \label{avphi}
\end{eqnarray}
In turn, the finite change of the  effective action $\Gamma_{M,\Psi}$ (as well as of $Z_{M_\Psi,\Psi}$)  under change of the gauge Fermion, $\Delta_f\Psi$, without using FDBRST transformations concept and as well the transformation rules  for BRST breaking term $M$  (\ref{reprmpsi+}) was firstly derived in \cite{Reshetnyak} (see Eq. (3.31)) from which the  linear in $\Delta_f\Psi$ and $\Delta_f M$ approximation looks as
\begin{eqnarray}\label{linGM}
&&  {\Delta}_f\Gamma_{M_\Psi,\Psi} \ = \frac{\delta\Gamma_{M_\Psi,\Psi}}{\delta\phi^A}
{\widehat F}^A\,{\Delta}_f\Psi({\widehat\phi})  -\ {\widehat
M}_A{\widehat F}^A {\Delta}_f\Psi({\widehat\phi})\ +\ {\Delta}_f M_\Psi({\widehat\phi},\phi^*) , \label{varGammaFlin}\\
&& \mathrm{where} \   {\widehat F}^A = -\frac{\delta}{\delta\phi^*_A}\ +\
(-1)^{\varepsilon_B(\varepsilon_A+1)} (\Gamma_{M,\Psi}^{''-1})^{BC}\Big(\frac{\delta_{
l}}{\delta\phi^C}\frac {\delta
\Gamma_{M,\Psi}}{\delta\phi^{*}_{A}}\Big)\frac{\delta_{ l}
}{\delta\phi^B}
\end{eqnarray}
and it was argued in \cite{llr1, lrr} to be non-vanishing on the extremals  $\Gamma_{M_\Psi,\Psi; A}=0$. However, a sufficient condition to vanish of ${\Delta}_f\Gamma_{M_\Psi,\Psi}\vert_{\Gamma_{M_\Psi,\Psi; A}=0}$ shown in  \cite{Reshetnyak},
\begin{equation}
  \label{BasRest}
{\Delta}_f M_\Psi({\widehat\phi},\phi^*)  = {\widehat M}_A{\widehat F}^A {\Delta}_f\Psi({\widehat\phi})\ ,
\end{equation}
 is always fulfilled and appears nothing else that average expectation value of the linear in ${\Delta}_f\Psi$  relation  (\ref{reprmpsi+}) with account for (\ref{solfeq}), presented as,
 ${\Delta}_f M_\Psi({\phi},\phi^*) = - (\imath/\hbar ) M_\Psi \overleftarrow{s}_e{\Delta}_f \Psi$.

 For YM theories the form of modified Ward identities and result of gauge dependence study  remain valid \cite{MRnew3} and  simplify because of the nilpotency of Slavnov generator $\overleftarrow{s}=\overleftarrow{s}_e$,   the jacobian of FDBRST transformations and solution of the compensation equation for change of the gauge  take the form
(\ref{aollambda}), (\ref{YMlambda}).

\section{Finite Field-Dependent BRST-antiBRST Transformation and its  Jacobian}\label{sec4}

The GFGF for irreducible  gauge theories with closed algebra within  BRST-antiBRST Lagrangian quantization
\cite{BLT1,BLT2} is given by,
 \begin{equation}
Z_F(J)=\textstyle\int d\phi\ \exp\left\{  \frac{i}{\hbar}\left[  S_{F}\left(  \phi\right)
+J_{A}\phi^{A}\right]  \right\}  \;. \label{z(jmm)}%
\end{equation}
 with  BRST-antiBRST-invariant quantum action
\begin{eqnarray}
\hspace{-0.5em}&& \hspace{-1em} S_{F}(\phi) = S_{0}\left(  A\right)  - 1/2  F_\xi \overleftarrow{s}_{a}\overleftarrow{s}^{a} =  S_{0}\left(  A\right)  +S_{\mathrm{gf}}\left(  A,B\right)
+S_{\mathrm{gh}}\left(  A,C\right)  +S_{\mathrm{add}}\left(  C\right) ,
\label{S(A,B,C)}%
\end{eqnarray}
determined on the total configuration space parameterized as above respectively by the classical, $Sp(2)$-duplet of ghost-antighost, Nakanishi-Lautrup  fields $\phi^A =( A^i, C^{\alpha a}, B^\alpha) $   and being the same as in FP method  under identification $(C^{\alpha 1},C^{\alpha 2})=(C^{\alpha},\overline{C}^{\alpha})$.
The quantities $S_0$, $F$ appear by classical gauge-invariant action  and admissible gauge-fixing Bosonic functional chosen here in quadratic approximation, in case of YM theory (with $A^i=A^{\mu n}(x)$ given on $D$-dimensional Minkowski space  for  $\eta_{\mu\nu}=diag(-,+,...,+)$  and   taking its values in the algebra Lie of  $SU(N)$ gauge group
\begin{eqnarray}\label{YMaction}
 &&  S_0 = - 1/4 \textstyle \int d^D x F_{\mu\nu}^nF^{\mu\nu{}n},\  \mathrm{for}\ F^{\mu\nu{}n} = \partial^{[\mu} A^{\nu] n}+ f^{nop}A^{\mu 0}A^{\nu p},\  n=1,...,N^2-1,
\\
&& \label{gfrxi}
F_\xi(  A,C)  =  - \frac{1}{2}\textstyle\int d^{D}x\ \left(  A_{\mu}%
^{m}A^{m\mu}-\xi/2\varepsilon_{ab}C^{ma}C^{mb}\right)
\end{eqnarray}
corresponding to $R_{\xi}$-family of the gauges (with $\chi_\xi(A,B)= \partial _{\mu }A^{\mu a}+\frac{%
\xi }{2}B^{a}=0$) within FP rules for YM theories.
The gauge-fixing term $S_{\mathrm{gf}}$, the ghost term $S_{\mathrm{gh}%
}$, and the interaction term $S_{\mathrm{add}}$, quartic in $C^{ma}$ in (\ref{S(A,B,C)})  (vanishing for Landau gauge $\xi=0$ and therefore for $S_F|_{\xi =0}$ coinciding with
FP BRST-invariant  action $S_{FP}(\phi)$) are determined by,
\begin{eqnarray}
\hspace{-0.7em}&&\hspace{-0.7em}\big(S_{\mathrm{gf}},\  S_{\mathrm{gh}}\big)   = \textstyle\int d^{D}x\Big(\left[  \left(  \partial^{\mu}A_{\mu
}^{m}\right)  + \xi/2 B^{m}\right]  B^{m}, \ \frac{1
}{2}\left(  \partial^{\mu}C^{ma}\right)  D_{\mu}^{mn}%
C^{nb}\varepsilon_{ab}  \Big) , \\
\hspace{-0.7em}&&\hspace{-0.7em}  S_{\mathrm{add}} = -\textstyle\frac{\xi}{48}\textstyle\int d^{D}x  f^{mnl}f^{lrs}%
C^{sa}C^{rc}C^{nb}C^{md}\varepsilon_{ab}\varepsilon_{cd}. %
\end{eqnarray}
The action (\ref{YMaction}) is invariant with respect to the infinitesimal gauge transformations
$ \delta A_{\mu}^{m}    = D_{\mu}^{mn}\zeta^{n}$ with arbitrary functions $\xi^\alpha \equiv \zeta^{n}$ ($\varepsilon_\alpha=0$) on  $R^{1,D-1}$,
whereas the infinitesimal BRST-antiBRST transformations, $\delta \phi^A =\phi^A \overleftarrow{s}^a\mu_{a}$,  for YM theories in terms of anticommutiong generators $\overleftarrow{s}^a: \overleftarrow{s}^a\overleftarrow{s}^b+\overleftarrow{s}^b\overleftarrow{s}^a=0$,
 \begin{eqnarray}
\hspace{-1.0em}&&\hspace{-0.7em} \Big(A_{\mu}^{m}, B^{m}\Big)\overleftarrow{s}^a  = \Big(D_{\mu}^{mn}C^{na} ,  {1}/{2}f^{nml}\hspace{-0.1em}\left[ \hspace{-0.1em} B^{l}C^{na}\hspace{-0.1em}+\hspace{-0.1em}({1}/{6})%
f^{lrs}C^{sb}C^{ra}C^{nc}\varepsilon_{cb}\hspace{-0.1em}\right]\hspace{-0.1em}\Big),\nonumber\\
\hspace{-1.0em}&&\hspace{-0.7em}  C^{ma} \overleftarrow{s}^b = \left(\hspace{-0.1em}  \varepsilon^{ab}B^{m}-({1}/{2})f^{mnl}%
C^{la}C^{nb}\right)   , \label{DCma}%
\end{eqnarray}
leave the action  $S_{F}$ and integrand $\mathcal{I}^F_{\phi}$ in $Z_F(0)= \int \mathcal{I}^F_{\phi}$ by invariant only in the $1$-st order in $\mu_{a}$.

To restore the total BRST-antiBRST invariance of $S_{F}$ and $\mathcal{I}^F_{\phi}$ in the whole orders in $\mu_{a}$ we introduced in \cite{MRnew}
finite transformations of  $\phi^{A}$ with a
doublet $\lambda_{a}$ of anticommuting parameters, $\lambda
_{a}\lambda_{b}+\lambda_{b}\lambda_{a}=0$,%
\begin{eqnarray}
\hspace{-1em}&& \hspace{-1.5em}\phi^{A}\rightarrow\phi^{\prime A}=\phi^{\prime
A}\left(  \phi|\lambda\right):  \phi^{\prime
}\left(  \phi|0\right)  =\phi, \; \textstyle\left[\hspace{-0.15em} %
\phi^{\prime
A} \frac{\overleftarrow{\partial}}{\partial\lambda_{a}}\hspace{-0.15em}\right]_{\lambda=0}=\phi^{A}\overleftarrow{s}^{a}\; \mathrm{and}%
\;  \textstyle \left[ \hspace{-0.15em} \phi^{\prime
A}\frac{\overleftarrow{\partial}}{\partial\lambda_{a}}%
\frac{\overleftarrow{\partial}}{\partial\lambda_{b}}\hspace{-0.15em}\right]\hspace{-0.1em}
=\hspace{-0.1em}\frac{1}{2}\varepsilon^{ab}\phi^{A}\overleftarrow{s}^{2}\label{Z}
\end{eqnarray}
as the solution of the functional equation
\begin{equation}\label{funceq}
  G\left(\phi'\right)  = G\left(\phi\right)\ \texttt{ if }\  s^a G\left(\phi\right) = 0
\end{equation}
for any regular  functional $G(\phi)$.
The general solution of (\ref{funceq}) permits to restore \emph{finite BRST-antiBRST transformations} in a unique way $\phi^{A} \to  \phi^{\prime A}$,
\begin{equation}
\hspace{-1.2em} \phi^{\prime A}\hspace{-0.1em} =\textstyle\hspace{-0.1em}  \phi^{A}\left(1\hspace{-0.1em}+   \overleftarrow{s}^{a}%
 \lambda_{a}\hspace{-0.1em}+\frac{1}{4} \hspace{-0.1em} \overleftarrow{s}^{2}
\lambda^{2}\right)\hspace{-0.1em} \equiv\hspace{-0.1em} \phi^{A} \exp\{\overleftarrow{s}^{a}%
 \lambda_{a} \hspace{-0.1em}\}, \label{finite1}%
\end{equation}
where a set of elements $\{g(\lambda_a)\}=\{\exp\{\overleftarrow{s}^{a}%
 \lambda_{a} \}\}$ forms Abelian two-parametric supergroup with odd generating elements $\lambda_{a}$.
 The BRST-antiBRST invariance of $\mathcal{I}^F_{\phi}$ means the validity of
 \begin{equation}\label{finbab}
   \mathcal{I}^F_{\phi g(\lambda_a)} = \mathcal{I}^F_{\phi}.
 \end{equation}
 where we have used the fact established in \cite{MRnew} that under global finite transformations, corresponding to
$\lambda_{a}=\mathrm{const}$, the integration measure remains invariant:
\begin{equation}
\mathrm{Sdet}\left(
\frac{\delta\phi\exp\{\overleftarrow{s}^{a}%
 \lambda_{a} \} }{\delta\phi}\right)  =1\ \ \mathrm{and}%
\ \ d\phi^{\prime}=d\phi. \label{constJ}%
\end{equation}
At the same time for  finite {field-dependent} transformations, we show in \cite{MRnew} that for
the particular case of  functionally
dependent parameters $\lambda_{a} =\Sigma \overleftarrow{s}_{a}$,   ($\lambda_{1}\overleftarrow{s}^{1}+\lambda_{2}\overleftarrow{s}^{2}= -\Sigma \overleftarrow{s}^{2}$)  with a certain
even-valued potential, $\Sigma=\Sigma\left(  \phi\right) $, which is
inspired by infinitesimal field-dependent BRST-antiBRST transformations with
the parameters
\begin{equation}
\mu_{a}=\frac{i}{2\hbar}\varepsilon_{ab}\left(  \Delta_f F\right)  _{,A}%
X^{Ab}=\frac{i}{2\hbar}\left(  \Delta_f F\right)\overleftarrow{s}_{a}  \,, \label{partic_case}%
\end{equation}
for which with accuracy up  to linear in $\Delta_f F$ terms the gauge independence of the integrand (therefore of the vacuum functional $Z_{F}(0)$) follows  $\mathcal{I}^F_{\phi g(\mu(\Delta_f F))}  = \mathcal{I}^{F+\Delta_f F}_{\phi} + o(\Delta_f F)$.  In  case of finite {field-dependent} transformations with group element $g(\Sigma \overleftarrow{s}_{a})$ {a set of which forms now non-Abelian 2-parametric supergroup}, the superdeterminant of the change of variables
takes the form %
\begin{eqnarray}
\hspace{-0.5em}&& \hspace{-1em} \mathrm{Sdet}\left(
\frac{\delta(\phi g(\Sigma\overleftarrow{s}_{a})}{\delta\phi}\right) = \left[  1-\frac{1}{2}%
\Sigma \overleftarrow{s}^{2}
\right]^{-2},  \  d\phi^{\prime} = \textstyle  d\phi\ \exp\left\{  \frac{i}{\hbar}\left[
i\hbar\,\mathrm{\ln}\left(  1-\frac{1}{2}\Sigma \overleftarrow{s}^2\right)
^{2}\right]  \right\}  \ . \label{superJ1}%
\end{eqnarray}
Again, the functionally dependent FDBRST-antiBRST transformations may be used due to $s_a$-exact form of the jacobian  (\ref{superJ1}) for the establishing of the gauge independence of the vacuum functional $Z_F(0)$ from the requirement of the BRST-antiBRST version  of the compensation equation under change of the gauge Boson, $F\to F+\Delta_fF$, validity:
\begin{eqnarray}
   &&  \mathcal{I}^F_{\phi g(\Sigma \overleftarrow{s}_a)} = \mathcal{I}^{F+\Delta_fF}_{\phi}  \Longleftrightarrow i\hbar\ \mathrm{\ln}\left(  1-\Sigma \overleftarrow{s}{}^2/2\right)  ^{2}%
\ =\ \left(  \Delta_f F \overleftarrow{s}{}^2 /2\right) ,\label{compbab}
\end{eqnarray}
whose solution for unknown Bosonic FD parameter $\Sigma(\phi)$, and therefore  for $Sp(2)$-doublet of $\lambda_a(\phi) = \Sigma \overleftarrow{s}_a $  with accuracy up to for ${s}_a$-exact terms looks as \cite{MRnew}:
\begin{eqnarray}
\hspace{-0.7em}&&\hspace{-0.5em}
\Sigma\left(  \phi|\Delta_f F\right)= -\textstyle 2\Delta_f F\left( \hspace{-0.1em} (\Delta_f F)\overleftarrow{s}{}^2\right)  ^{-1}\left[ \hspace{-0.2em} \exp\left(\hspace{-0.1em}
-\frac{1}{4i\hbar}(\Delta_f F) \overleftarrow{s}{}^2\right) \hspace{-0.1em} -1\right]   . \label{Sigma-Fsol1}
\end{eqnarray}
And visa-verse having considered the equation (\ref{compbab}) for unknown $\Delta_f F$ with given $\Sigma$ we obtain
\begin{eqnarray}\hspace{-0.7em}&&\hspace{-0.5em}
\Delta_f F\left(  \phi\right)  =\textstyle- 2i\hbar\ \Sigma \left(
\Sigma \overleftarrow{s}{}^2 \right)  ^{-1}\ln\left(  1-%
\Sigma  \overleftarrow{s}{}^2 /2\right)  ^{2}\ . \label{Lambda-Fsol}%
\end{eqnarray}
 Thus, the
field-dependent transformations  with the parameters $\lambda_{a}=\Sigma\overleftarrow{s}_a $
amount to a precise change of the gauge-fixing functional.
E.g. to relate $Z_{F_\xi}(J)$ with $Z_{F_{\xi+\Delta \xi }}(J)$ in $R_\xi$- family of the gauges
we should to fulfill FFDBRST-antiBRST transformations with parameters, $\lambda_{a}=\lambda_{a}(\xi)$
\begin{eqnarray}
\hspace{-1.4em}&&\hspace{-1.3em}\lambda_{a} \textstyle =\frac{\Delta\xi}{4i\hbar}\varepsilon_{ab}\int
d^{D}x\  B^{n}C^{nb}\sum\limits_{n=0}^{\infty}\hspace{-0.2em}\frac{1}{\left(  n+1\right)  !}
\hspace{-0.2em}\left[\hspace{-0.2em}
\frac
{\Delta\xi}{4i\hbar}\int d^{D}y \hspace{-0.15em}\left(  \hspace{-0.2em}B^{u}B^{u}-\frac{1}{24}%
f^{uwt}f^{trs}C^{sc}C^{rp}C^{wd}C^{uq}\varepsilon_{cd}\varepsilon
_{pq}\right)  \hspace{-0.2em}\right]  ^{n}\hspace{-0.2em}. \label{lamaxi}%
\end{eqnarray}
Being base on (\ref{Sigma-Fsol1}) the  modified Ward identity for $Z_F(J)$ depending on FD parameters $\lambda_a(\phi|\Delta_fF)$, following from it usual Ward identities
for constant $\lambda_a$ and gauge dependence problem under finite change of the gauge \cite{MRnew, MRnew3} in terms of the notations similar to one in  (\ref{mWIbvbr}) :
\begin{eqnarray}
\hspace{-1.3em}&&\hspace{-1.4em} \textstyle\left\langle \left\{  1+\frac{i}{\hbar}J_{A}\left[ \phi^A\overleftarrow{s}{}^a \lambda_{a}%
(\Sigma)+\frac{1}{4}\phi^A\overleftarrow{s}{}^2 \lambda^{2}(\Sigma)\right]-\frac{1}{4}\left(
\frac{i}{\hbar}\right)  {}^{2}\varepsilon_{ab}   J_{A}\phi^A\overleftarrow{s}{}^a  J_{B}\phi^B\overleftarrow{s}{}^b %
\lambda^{2}(\Sigma)\right\} \right. \label{mWIclalg}\\
&& \left.\times\left(  1-\textstyle\frac{1}{2}\Sigma\overleftarrow
{s}^{2}\right)  {}^{-2}\right\rangle _{F,J} =1, \qquad  J_{A}\left\langle \phi^A\overleftarrow{s}{}^a \right\rangle _{F,J}=0, \label{WIusbab}\\
\hspace{-1.3em}&&\hspace{-1.4em} \Delta Z_{F}(J)\textstyle = \hspace{-0.15em}\frac{i}{\hbar}Z_{F}\left\langle %
J_{A}\left[\hspace{-0.15em} \phi^A\overleftarrow{s}{}^a  \hat{\lambda}_{a}\hspace{-0.15em}   +\hspace{-0.15em}\frac{1}%
{4}\phi^A\overleftarrow{s}{}^2 \hat{\lambda}^2 \hspace{-0.15em} \right]\hspace{-0.15em} -  \hspace{-0.15em} (-1)^{\varepsilon_{B}}\hspace{-0.15em}\left(  \frac{i}{4\hbar}\right)  %
J_{B}J_{A}\left(\hspace{-0.15em}  \phi^A\overleftarrow{s}{}^a \phi^B\overleftarrow{s}{}^b \hspace{-0.15em}\right)  \varepsilon_{ab}\hat{ \lambda}^{2} \hspace{-0.15em}\right\rangle _{F,J} \label{GDInew1}%
\end{eqnarray}
for the notations  $\hat{\lambda}_{a} \equiv \lambda_{a}\left( \phi| -\Delta_f{F}\right)$.

\section{(Refined) Gribov--Zwanziger Theory in BRST, BRST-anti\-BRST Formulations in Many Parametric Family of  Gauges}\label{sec5}

GZ theory is determined by the GZ
action $S_{GZ}(\phi)$ on the same configuration space as for YM theory, given in the Landau gauge $\chi^m(A)=\partial_\mu  A^{\mu  m}=0$ (with using of the Minkowsky space-time notations rather formally because of GZ theory is determined only in Euclidian space $\mathbb{R}^D$)
\begin{eqnarray}  \label{GZact}
S_{GZ}(\phi)=S_{FP}(\phi) + M_0(A) ,\  M_0(A)= \gamma^2\big(f^{mrl}%
A_{\mu}^{r}   {K ^{mn}}^{-1} f^{nsl}A^{\mu{}s } + D(N^2-1)\big)
\end{eqnarray}
with the additive non-local BRST-non-invariant  with respect to BRST transformations $\bigl(A^{\mu m}$,  $C^{m}$,  $\overline{C}{}^{m}$,  $B^m\bigr)\overleftarrow{s}$ = $\bigl( D^{\mu mn}C^n, \frac{1}{2}f^{mno}C^{n}C^{o},
B^{m},\ 0\bigr)$ term:
\begin{equation}  \label{sM}
M_0\overleftarrow{s}  =\gamma^2f^{mrl}f^{lse}\bigl[2D^{rq}_{\mu}C^q(K^{-1})^{ms}-
f^{gpn}A^r_{\mu}(K^{-1})^{mg}K^{pq}C^q(K^{-1})^{ns}\bigr]A^{e\mu}\ \neq\ 0,
\end{equation}
 known as the GZ horizon functional,
implying an inclusion of the Gribov horizon  \cite{Gribov} in terms of the FP operator $(K)^{mn} = \partial_\mu D^{%
\mu{}mn}$ for $  (K^{-1})^{mo}(K)^{on}=\delta^{mn}$
and the so-called thermodynamic (Gribov mass)
parameter~$\gamma$, introduced
in a self-consistent way by the gap equation \cite{Zwanziger}.   The idea to improve the GZ theory is
due to the facts that,  first, it fails to
eliminate all Gribov's copies, and, second, a non-zero value for
the Gribov parameter $\gamma$ is a manifestation of nontrivial
properties of the vacuum \cite{0806.0348} of the theory. The latter
means that there exist additional reasons for
non-perturbative effects, which can be encoded in a set of
dimension-2 condensate, $\langle A^{\mu a}A_\mu^a \rangle $, in
the case of a non-local GZ action with the
YM gauge fields $A^{\mu a}$ only\footnote{As well as
in a similar set of dimension-2 condensates, $%
\langle A^{\mu m}A_\mu^m \rangle$, $\langle \bar{\varphi}^{\mu{}
mn}\varphi_\mu^{mn} \rangle{-}\langle \bar{\omega}^{\mu{}
mn}\omega_\mu^{mn} \rangle$, for a local GZ action \cite{0806.0348},
$S_{GZ}(\phi, \hat{\phi})$ with an equivalent local representation for
the horizon functional in terms of the functional $S_\gamma$,
defined in an extended configuration space with auxiliary variables
$\phi^{\bar{A}}$ described in \cite{Zwanziger1},\cite{Reshetnyak}.}
\begin{eqnarray}
&& S_{GZ}(\phi) \rightarrow S_{RGZ}(\phi)=S_{GZ}+\textstyle\frac{m^2}{2} {A^m_{\mu}}{%
A^{\mu{}m}} .  \label{RGZn}
\end{eqnarray}
To determine GZ and RGZ models in any  gauges  in a gauge independence  manner compatible with (\ref{GDInewb})
let us consider a family of linear gauges given by the equation
\begin{eqnarray}
\chi ^{m}(A,B) &=&\Lambda _{\mu }(\partial ,\alpha ,\beta ,n)A^{\mu m}+\frac{%
\xi }{2}B^{m}=0 \  \mathrm{ with }\ \Lambda _{\mu }(\partial ,\alpha ,\beta ,n) = \alpha \partial _{\mu }+\beta
\frac{\kappa_{\mu\nu}}{n^2}n^{\nu } . \label{gengauge}
\end{eqnarray}%
Here, we have 3 real, $\alpha ,\beta ,\xi $, and 1 vector, $n^{\mu }$,
gauge parameters.

Particular cases of $R_\xi$-gauges and \emph{generalized  Coulomb
gauges}  gauges can be obtained from the general
many-parameter family under the choices
\begin{eqnarray}
&& (\alpha;\beta) = (1;0) \rightarrow \ R_{\xi -}\mathrm{gauges};
\label{lingage1}\  \
(\beta,\xi) =( - \alpha,0) , \kappa_{\mu\nu} = n^{\rho }\partial _{\rho}\eta_{\mu\nu},  n^{2}<0 ,
\end{eqnarray}%
The Landau and Feynman  gauges are obtained from the first family for the
respective choices $\xi =0; 1$, whereas  the  Coulomb, $\chi_{C}^{m}(A,B)=\partial _{i}A^{i m}=0$ for $\mu
=(0,i)$  from $n^{\mu }=(1,0,...,0)$.

The FP action, GZ horizon functional $M_g(\phi)$ and, therefore GZ and RGZ model in any gauges, including ones  from the set of (\ref{gengauge}) starting from ones in the Landau (or Coulomb \cite{GZ2}, where horizon functional has the same form (\ref{GZact}) but for $(K)_C^{ab} = \partial_i D^{%
i{}ab}$ and $D-1$ instead of $D$) gauge is determined by (\ref{reprmpsi+}) with help  of FDBRST transformations with odd parameter $\hat{\Lambda}$ from (\ref{YMlambda}) with $\overleftarrow{s}$ defined before (\ref{sM}):
\begin{eqnarray}\label{FPagen}
 &\hspace{-1.5em}& \hspace{-1em} S_{FP}({\phi,\alpha,\beta, n^\mu, \xi})=  S_0 + \Psi_g(\phi)\overleftarrow{s},\ \ \mathrm{for}\ \ K_g^{mn}=\Lambda^\mu(\partial,\alpha,\beta,n) D_\mu^{mn}, \\
 &\hspace{-1.5em}& \hspace{-1em} M_g(\phi) = M_0(A)\exp\{\overleftarrow{s}\hat{\Lambda} \} = M_0(A)\hspace{-0.2em}\left(1+ \overleftarrow{s}{\Delta }_f\Psi \big\{\bigl({\Delta}_f\Psi\big)\overleftarrow{s}\big\}%
^{-1}\Big[\hspace{-0.1em}\exp \hspace{-0.2em}\left\{\textstyle \hspace{-0.2em} -\frac{\imath}{\hbar }\bigl({\Delta}_f\Psi\big)\hspace{-0.1em}\overleftarrow{s}\hspace{-0.2em}\right\} \hspace{-0.1em}-1\hspace{-0.1em}\Big]\hspace{-0.2em}\right),\label{Mgen}\\
 &\hspace{-1.5em}& \hspace{-1em} \textstyle\frac{m^2}{2} {A^m_{\mu}}{A^m_{\mu}}\exp\{\overleftarrow{s}\hat{\Lambda} \} =  \frac{m^2}{2}{A^m_{\mu}}\hspace{-0.2em}\left( {A^m_{\mu}}\hspace{-0.1em} + \hspace{-0.1em}\partial^{\mu}C^m{\Delta }_f\Psi \big\{\bigl({\Delta}_f\Psi\big)\overleftarrow{s}\big\}%
^{-1}\big[\exp \hspace{-0.2em}\left\{\textstyle -\frac{\imath}{\hbar }\bigl({\Delta}_f\Psi\big)\hspace{-0.1em}\overleftarrow{s}\hspace{-0.1em}\right\} \hspace{-0.1em}-1\big] \hspace{-0.2em}\right),\label{comptr}
\end{eqnarray}
for $\Psi_g(\phi) = {\bar C}^m \chi ^{m}(A,B)$ and where
\begin{eqnarray}
 &\hspace{-1.5em}& \hspace{-1em}\Delta_f \Psi = \Psi_g - \Psi_0 = \textstyle\bar{C}{}^{m}\big(\{(\alpha -1)\partial _{\mu }+\beta
\frac{\kappa_{\mu\nu}}{n^2}n^{\nu }\}A^{\mu m}+\frac{\xi }{2}B^{m}%
\big), \label{deltafp}\\
 &\hspace{-1.5em}& \hspace{-1em}{\Delta}_f\Psi \overleftarrow{s}  = \hspace{-0.2em} \textstyle\Big\{\hspace{-0.2em}B^{m}\Big(\hspace{-0.2em}\{(\alpha
-1)\partial _{\mu }\hspace{-0.1em}+\hspace{-0.2em}\beta \frac{\kappa_{\mu\nu}}{n^2}n^{\nu }\}A^{\mu m}+\frac{\xi }{2}B^{m}\hspace{-0.1em}\Big)  \hspace{-0.1em} +\hspace{-0.2em}\bar{C}{}^{m}\big((\alpha -1)\partial
_{\mu }\hspace{-0.1em}+\hspace{-0.1em}\beta \frac{\kappa_{\mu\nu}}{n^2}n^{\nu }\big)D^{\mu
mn}C^{n}\hspace{-0.2em}\Big\}.  \label{sdeltapsiYM}
\end{eqnarray}
The GZ: $S_{g;GZ}=S_{FP}({\alpha,\beta, n^\mu, \xi})+ M_g(\phi)$,  and RGZ: $S_{g;RGZ}=S_{g;GZ}+ \frac{m^2}{2} {A^m_{\mu}}{A^m_{\mu}}\exp\{\overleftarrow{s}\hat{\Lambda} \}$,  actions in any from $\chi ^{m}(A,B)$-set
of the gauges present one from the main results in this
section\footnote{These results call for a verification of the fact that
$M_g$ actually selects the first Gribov region for $A^{\mu{}a}$
in any  $\chi ^{m}(A,B)$-gauge, since extracting this region
by means of the functional $M_0(A)$ has been determined
non-perturbatively \cite{Zwanziger1}, whereas a corresponding
explicit and rigorous proof, e.g., for an $R_\xi$-gauge  $M(A,\xi)$
to provide a restriction for $A^{\mu{}a}$ in the first
Gribov region, $%
\Omega(\xi)$:
$ = \Big\{A^{\mu{}a}\big\vert \chi ^{a}(A,B)\big\vert_{\alpha
=1,\beta =0}=0, K^{ab}(\xi)\geq 0 \Big\}$,
has not been presented in the literature.}
Considering the generalization of GZ and RGZ theories within BRST-antiBRST quantization, note because of  the gauge-fixing functional $F_0$ (\ref{gfrxi}) corresponds
to the Landau gauge, we introduce the GZ horizon
functional in the same manner as in \cite{Zwanziger}
for the FP procedure in the Euclidian space coinciding with $M_{F_0}= M_0(A)$ (\ref{GZact}) as well as GZ  action appears by the same (\ref{GZact}.
We determine the GZ theory
in any $F_\xi$ gauges ($R_\xi$-gauges) in a way
compatible with the gauge-independence of the
generating functional of Green's functions in $F_0$,
where Gribov horizon in the gauge $F_\xi$
should be determined as
 \begin{eqnarray}
\hspace{-0.9em}&&\hspace{-0.5em} M_{F_\xi}    = M_{F_0}\Big(1+\textstyle\frac{1}{2i\hbar}\left(  \overleftarrow{s}^{a}\right)  \left(  \Delta
F_{\xi}\overleftarrow{s}_{a}  \right) \sum_{n=0}^{\infty}\frac{1}{\left(
n+1\right)  !} \left(-  \textstyle\frac{1}{4i\hbar}\Delta F_{ \xi}\overleftarrow{s}^2
\right)  ^{n}  -\textstyle\frac{1}{16\hbar^{2}}\left(  \overleftarrow{s}^{2}\right)  \left(  \Delta F_{ \xi}  \right)  ^{2}\nonumber\\
\hspace{-0.9em}&&\hspace{-0.0em} \times \left[  \textstyle\sum_{n=0}^{\infty}\frac{1}{\left(
n+1\right)  !}\left( - \frac{1}{4i\hbar}\Delta F_{ \xi}\overleftarrow{s}^2
\right)  ^{n}\right]  ^{2}\Big) , \label{gribovHxi}%
\end{eqnarray}
where $\Delta F_{ \xi}$ is readily determined with account taken
of (\ref{DCma}), (\ref{lamaxi}); see for details  \cite{MRnew}.
The construction of the Gribov horizon functional $M_{F_\xi}$ ($h_{\xi}$ in \cite{MRnew})
in the gauge $F_\xi$, starting from $M_0$ in the gauge $F_0$,
may be considered as a generalization of the result \cite{LL2}
obtained in the BRST setting of the problem. In  turn, the RGZ model in any from ${F_\xi}$-gauge are readily constructed as, $S_{RGZ,\xi}$ with account for
  \begin{eqnarray}
  \textstyle\frac{m^2}{2} {A^m_{\mu}}{A^m_{\mu}}\exp\{\overleftarrow{s}{}^a\lambda_{a}(\xi)\} =  \frac{m^2}{2}\Big[{A^m_{\mu}}\hspace{-0.2em}\left( {A^m_{\mu}}\hspace{-0.1em} + \hspace{-0.1em}\partial^{\mu}C^{ma}\lambda_{a}(\xi) \right) + \frac{1}{4}A^m_{\mu}A^{\mu m}\overleftarrow{s}{}^2\lambda^2(\xi) \Big], \label{comptrbab}
\end{eqnarray}
which is differed by the last term proportional to $\lambda^2(\xi)$ from BRST transformed composite field (\ref{comptr}).

\section{Modified Faddeev-Popov Rules for  Gauge Theory with Gauge Group}\label{sec6}

Starting from the Gribov anzats for YM theory with the functional $\mathcal{V}(\partial_\mu\partial^\mu)$ in the Eq. (31) \cite{Gribov} which restricts the integration in the path integral, $Z_F(0)$ (\ref{z(jmm)}) in BRST-antiBRST quatization (or $Z_\Psi(0)$ within FP method with $S_{FP}$ instead of $S_F$)  only to the first Gribov region $C_0$
we suppose that it may be presented as $\Theta$-function: $\mathcal{V}(\partial_\mu\partial^\mu) = \Theta(\partial_\mu\partial^\mu) = \Theta(1-\sigma(\lambda_0(A)))$, where a quantity $\lambda_0(A)$ appears by the least real part of positive proper eigen-value of the FP operator $K^{mn}(A)$ in a gauge $\chi^n=0$:
$ 0 \geq  \mathrm{Re}\lambda_0^n \geq \mathrm{Re}\lambda_1^n \geq  \ldots \geq \mathrm{Re}\lambda_k^n \geq \ldots $ in the spectrum problem on,
\begin{equation}\label{FDev}
  K^{mn}(A) u^n_k = \delta^{mn}\lambda_k^n(A) u^n_k,\ \mathrm{for}\ K^{mn}(A) = ({\delta\chi^m})/({\delta A^{\mu o}})D^{\mu on},\  k \in \mathbb{Z}.
\end{equation}
We determine the GFGF $\mathcal{Z}\Psi(J)$ with restricted region of the integration (where $\mathrm{Det}\| K^{mn}(A)\|$ $> 0$ everywhere) for the gauge Lie algebra $g=su(N)$ without Gribov's copies as,
\begin{eqnarray}\label{FPtrick}
\mathcal{Z}_\Psi(J)&   = &  \int dA \Theta[1-\sigma(\lambda_0(A))] \delta\big(\chi(A)\big) \mathrm{Det} K(A)\exp\big\{\textstyle\frac{\imath}{\hbar}(S_0(A)+ jA) \big\}  \end{eqnarray}\vspace{-1ex}
\begin{eqnarray}
\phantom{\mathcal{Z}_\Psi(J)}&   = &  \int dA  \delta\big(\chi(A)\big)\det \left\{\Theta^{\frac{1}{\dim g}}[1-\sigma(\lambda_0(A))]K(A)\right\}\exp\big\{\frac{\textstyle\imath}{\hbar}(S_0(A)+ jA) \big\}  \nonumber \\
&=& \int d\phi\exp\big\{\textstyle\frac{\imath}{\hbar}(S_0(A)+ \bar{C}K'C + \chi(A)B +  jA) \big\},
\label{FPtrick1}\end{eqnarray}
with    $K^{\prime mn }= \Theta^{\frac{1}{\dim g}}[1-\sigma(\lambda_0(A))]K^{mn}(A)$, being by modified Faddeev-Popov operator in the Lagrangian formalism, for $\dim g=N^2-1$.
Of course, it is a problem to solve the  spectrum problem (\ref{FDev}) and to construct the functional $\sigma(\lambda_0(A))$ but corresponding results for some gauge group exist.

The classical action, $S_0(A)$ is still invariant with respect to
  by the restricted to the region $C_0$ infinitesimal \emph{modified gauge transformations} with \emph{modified generators of gauge transformations}:
  \begin{eqnarray}\label{mgtr}
   && \delta_m A^{\mu n}(x) = \Theta^{\frac{1}{\dim g}}[1-\sigma(\lambda_0(A))]D^{\mu no}(x) \zeta^o(x),\\
    && \mathrm{with}\    \mathcal{R}^{\mu m o}(x,y) = \Theta^{\frac{1}{\dim g}}[1-\sigma(\lambda_0(A))]D^{\mu m o}(x)\delta(x-y) \nonumber.
 \end{eqnarray}
  The integrand $\mathfrak{I}^\Psi_{\phi}$  in $\mathcal{Z}_\Psi(J) = \int \mathfrak{I}^\Psi_{\phi} \exp \{\frac{\imath}{\hbar}J\phi\}$ is invariant with respect to \emph{modified BRST transformations}:
  \begin{equation}\label{mBRSTtrans}
   \delta_B \bigl(A^{\mu m},  C^{m},  \overline{C}{}^{m},  B^m\bigr) = \Theta^{\frac{1}{\dim g}}[1-\sigma(\lambda_0(A))]\bigl( D^{\mu mn}C^n, \frac{1}{2}f^{mno}C^{n}C^{o},
B^{m},\ 0\bigr)\Lambda,
  \end{equation}
where we have taken into account for the calculation of the jacobian of the change of variables in $\mathcal{Z}_\Psi(J)$ that the terms $(\delta \Theta(...))/(\delta A^i)$ should be proportional to $\delta(0)$ and within appropriate choice of the regularization should vanish.
We see that the gauge independence property for the vacuum functional should be follow as well as a consistency of the unitarity problem due to non-appearance of non-physical degrees of freedom as for the GZ model and suppose to continue this research in a forthcoming paper.

\section{Conclusion}\label{concl}

We have reviewed the results of our research
devoted to finite FDBRST transformations
in the BV formalism and calculated the Jacobian
of a change of variables, used afterwards
to obtain a new form of the Ward identities
for the generating functionals of Green's functions.
For these functionals, we study the issue
of gauge dependence, and this enables us
to solve the consistency problem of
an introduction of (soft) BRST symmetry breaking
terms in the BV method.
We have also proposed the  concept of finite
BRST-antiBRST and FFDBRST-antiBRST
transformations for Yang--Mills theories in the $\mathrm{Sp}(2)$-covariant
Lagrangian quantization.
The Jacobian of a change of variables corresponding
to FFDBRST-antiBRST transformations
with functionally-dependent parameters is
calculated in a precise manner.
It has been established that quantum YM actions
in different gauges are related to each other
by FFDBRST-antiBRST transformations with functionally
dependent parameters obtained as solutions
of the compensation equation.
A new Ward identity and the gauge dependence
problem for finite changes of the gauge
for the  generating functional of Green's functions
have been obtained and studied.
The Gribov--Zwanziger theory and a refined
Gribov--Zwanziger theory in BRST and
in BRST-antiBRST descriptions for a many parametric
family of linear gauges (explicitly including
the covariant and Coulomb gauges), starting from
$M_0$ in the Landau gauge, are suggested in
a way consistent with the gauge independence
of the respective $S$-matrices as a consequence
of BRST(antiBRST) symmetry breaking.
A  modification is proposed for the Faddeev--Popov rules
to a definition that involves such a gauge
in the path integral and such BRST transformations
that are free from the Gribov copies and do not
excite the longitudinal degrees of freedom.

\section*{Acknowledgement}
The author thanks  V.A. Rubakov and  the Organizing Committee  for
kind hospitality. He is also grateful to
P.Yu. Moshin, D. Bykov, D. Francia, and to the participants
of the International Seminar ``QUARKS 2014'' for useful discussions.
The study has been supported by the LRSS grant under Project No. 88.2014.2.

\end{document}